\documentclass[pra,aps,floatfix,amsmath,superscriptaddress,twocolumn]{revtex4}

\usepackage{amssymb}
\usepackage{graphicx}
\usepackage{graphics}
\usepackage{amsmath}
\usepackage{amsthm}
\usepackage{color}
\usepackage{dsfont}

\usepackage{mathtools}
\def\ra{\rangle}
\def\la{\langle}

\newcommand{\bea}{\begin{eqnarray}}
\newcommand{\eea}{\end{eqnarray}}
\newcommand{\be}{\begin{equation}}
\newcommand{\ee}{\end{equation}}
\newcommand{\ba}{\begin{equation}\begin{aligned}}
\newcommand{\ea}{\end{aligned}\end{equation}}
\newcommand{\rank}{\text{Rank}}

\theoremstyle{remark}
\newtheorem{remark}{Remark}

\def\be{\begin{equation}}
\def\ee{\end{equation}}

\newcommand{\mH}{\mathcal{H}}

\newcommand{\mK}{\mathcal{K}}

\newcommand{\bv}{\mathbf{v}}
\newcommand{\bc}{\mathbf{c}}

\newcommand{\spa}{\text{span}}

\newcommand{\mS}{\mathcal{S}}

\newcommand{\lr}{\rangle\langle}

\newcommand{\tr}{{\rm Tr}}

\newcommand{\mbb}[1]{\mathbb{#1}}

\def\>{\rangle}
\def\<{\langle}

\begin{document}
	
	\preprint{APS/123-QED}
	\title{Monogamy of the entanglement of formation}% Force line breaks with \\

	\author{Yu Guo}
	\email{guoyu3@aliyun.com}
	\affiliation{Institute of Quantum Information Science, School of Mathematics and Statistics, Shanxi Datong University, Datong, Shanxi 037009, China}
	
	\author{Gilad Gour}
	\email{giladgour@gmail.com}
	\affiliation{ Department of Mathematics and Statistics, Institute for Quantum Science and Technology, 
		University of Calgary, Calgary, Alberta T2N 1N4, Canada}

	\begin{abstract}
	We show that any measure of entanglement that on pure bipartite states is given by a strictly concave function of the reduced density matrix is monogamous on pure tripartite states. 
	This includes the important class of bipartite measures of entanglement that reduce to the (von Neumann) entropy of entanglement. Moreover, we show that the convex roof extension of such measures (e.g., entanglement of formation) are monogamous also on \emph{mixed} tripartite states. To prove our results, we use the definition of monogamy without inequalities, recently put forward~[Gour and Guo, Quantum \textbf{2}, 81 (2018)]. Our results promote the concept that monogamy of entanglement is a property of quantum entanglement and not an attribute of some particular measures of entanglement.
	\end{abstract}
	
	%\pacs{03.67.Mn, 03.65.Ud.}% PACS, the Physics and Astronomy
	% Classification Scheme.
	%\keywords{Suggested keywords}%Use showkeys class option if keyword
	%display desired
	\maketitle
	%\end{CJK*}
	%\tableofcontents

	Quantum entanglement is one of the most counter-intuitive phenomena of quantum theory. In the early days of quantum mechanics it was recognized as ``the characteristic
	trait of quantum mechanics the one that enforces its entire departure from classical lines of thought.''~\cite{Schrodinger}. On the other hand, in recent years it was identified as the
	key resource of many
	quantum information processing tasks~\cite{Nielsenbook,Wildebook,Watrousbook}. One of its key features is that it
	cannot be freely shared among many parties, unlike
	classical correlations. This is the so-called monogamy
	law~\cite{Terhal2004} and is one of the fundamental traits of entanglement
	and of quantum mechanics itself~\cite{Adesso2006}.
	This shareability relation has been explored extensively~\cite{Coffman,Zhuxuena2014pra,Zhuxuena2015pra,Osborne,Ouyongcheng,
		Ouyongcheng2007pra2,Luo,Kim2009,Choi,Kumar,Bai,Oliveira2014pra,
		Koashi,Kim2016pra,Luo2016pra,Kim2010jpa,Cornelio,Song,Dhar,
		Chengshuming,Allen,Deng,Camalet,Karczewski,Guo} ever since Coffman, Kundu,
	and Wootters proved the first quantitative monogamy relation~\cite{Coffman} for three-qubit
	states.

	An important question in the study of monogamy of entanglement is to determine
	whether or not a given entanglement measure is monogamous.
	For multiqubit systems, almost all the known entanglement 
	measures are monogamous. These include the entanglement of formation~\cite{Bennett1996pra}, concurrence~\cite{Rungta,Rungta2003pra}, tangle~\cite{Rungta2003pra},
	negativity~\cite{Zyczkowski,VidalWerner}, convex-roof extended negativity~\cite{Lee},
	Tsallis-$q$ entropy of entanglement~\cite{Kim2010pra}, R\'{e}nyi-$\alpha$ entropy of entanglement~\cite{Kim2010jpa,Gour},
	squashed entanglement~\cite{Christandl2004jmp} and one-way distillable entanglement~\cite{Bennett1996prl,Bennett1996pra,Koashi}, 
	which are proved to be monogamous~\cite{Coffman,Zhuxuena2014pra,Zhuxuena2015pra,Osborne,Ouyongcheng,
		Ouyongcheng2007pra2,Luo,Kim2009,Choi,Kumar,Bai,Oliveira2014pra,
		Koashi,Kim2016pra,Luo2016pra,Kim2010jpa,Cornelio,Song}.
	However, for higher-dimensional systems much less is known about
	this shareability~\cite{Hehuan,Renxijun,Kim}. 
	So far, in addition to the one-way distillable entanglement~\cite[Theorem 6]{Koashi}
	and squashed entanglement~\cite[Theorem 8]{Koashi}, we know only
	that the G-concurrence~\cite{Gour2005} is monogamous in all finite dimensions~\cite{GG}.
	The latter are monogamous according to the definition given 
	in Ref.~\cite[Definition 1]{GG} (also see Eq.~\eqref{cond} below). 
	That is, the monogamy of other measures of entanglement in higher-dimensional systems still remain
	unknown even for well-known operational measures, such as the entanglement of formation.

	Let $\mH^{A}\otimes\mH^B\otimes\mH^C\equiv\mH^{ABC}$ be a tripartite Hilbert space with finite dimension, where $A,B,C$ are three subsystems of a composite quantum system, 
	and $\mS(\mH^{ABC})\equiv\mS^{ABC}$ be the set of density matrices acting 
	on $\mH^{ABC}$.
	Recall that the original monogamy relation of entanglement 
	measure $E$ is quantitatively displayed as an inequality of the following form:
	\be\label{basic}
	E(\rho^{A|BC}) \geq E( \rho^{AB}) +E( \rho^{AC}),
	\ee
	where the vertical bar indicates the bipartite split 
	across which  the (bipartite) entanglement is measured.
	However, Eq.~\eqref{basic} is not valid for many entanglement measures~\cite{Coffman,Kumar,Zhuxuena2014pra,Zhuxuena2015pra,Osborne,Kim2009,
		Luo,Bai,Ouyongcheng2007pra2,Choi}. This may give the impression that monogamy of entanglement is not a property of entanglement itself but of the function that is used to quantify it.

Moreover, in Ref.~\cite{Lan16} the problem of faithfulness versus monogamy was raised by showing that many measures of entanglement, such as the entanglement of formation, cannot satisfy any relation of the form
\be\label{basic2}
	E( \rho^{A|BC}) \geq f [E( \rho^{AB}) ,E(\rho^{AC})],
	\ee
where $f:\mbb{R}_{+}\times\mbb{R}_+\to\mbb{R}_+$ is some fixed function that is independent on the dimension of the underlying Hilbert space, and that it is continuous, and satisfies $f(x,y)\geq \max\{x,y\}$ with strict inequality at some ranges of $x$ and $y$. Despite this remarkable result, we show here that the entanglement of formation, and many other measures of entanglement that are defined in terms of convex roof extensions, are monogamous according to the definition recently put forward in Ref.~\cite{GG}.

According to the definition in Ref.~\cite{GG} of monogamy (without inequalities), 
	a measure of entanglement $E$ is monogamous if for any $\rho^{ABC}\in\mS^{ABC}$
	that satisfies the \textit{disentangling condition}, i.e.,
	\be\label{cond}
	E( \rho^{A|BC}) =E( \rho^{AB}),
	\ee
	we have that $E(\rho^{AC})=0$. With respect to this definition, if the entanglement between system $A$ and the composite system $BC$ is as much as the entanglement that system $A$ shares \emph{just} with $B$, then there is no entanglement left for $A$ to share just with $C$. 
	
	Clearly, this definition captures the spirit of monogamy of entanglement, and perhaps not surprisingly,
	can yield a family of monogamy relations similar to~\eqref{basic} by replacing $E$ with $E^\alpha$ for 
	some $\alpha>0$~\cite{GG}. More precisely, a continuous measure $E$ is monogamous according to this definition if and only if
	there exists $0<\alpha<\infty$ such that
	\be\label{power}
	E^\alpha( \rho^{A|BC}) \geq E^\alpha( \rho^{AB}) +E^\alpha( \rho^{AC}),	
	\ee
	for all $\rho^{ABC}$ acting on the state space 
	$\mH^{ABC}$ 
	with fixed $\dim\mH^{ABC}=d<\infty$ (see Theorem 1 in Ref.~\cite{GG}). Note that~\eqref{power} can be expressed in a similar form as in~\eqref{basic2} with $f(x,y)=(x^\alpha+y^{\alpha})^{1/\alpha}$. 
	However, we stress here that~\eqref{power} is not a special case of~\eqref{basic2} since the exponent factor $\alpha$ depends on the underlying dimension of the Hilbert space. Note that there is no \textit{a priori} physical reason to assume that the exponent factor is universal and independent on the dimension. 
	
	In this paper, we show that almost
	all entanglement monotones are monogamous on pure tripartite states, and furthermore, those that are based on convex roof extension [e.g., entanglement of formation, see also Eq.~\eqref{eofmin}],
	are also monogamous on mixed tripartite states according to our definition in terms of Eq.~\eqref{cond} [or equivalently Eq.~\eqref{power}]. 
		Our results indicate that monogamy is indeed a property of entanglement and not a consequence of a particular measure of entanglement.

	A function $E: \mS^{AB}\to\mbb{R}_{+}$ is called a 
	measure of entanglement if \textbf{(1)} $E(\sigma^{AB})=0$ for any 
	separable density matrix $\sigma^{AB}\in\mS^{AB}$, and \textbf{(2)} 
	$E$ behaves monotonically under local operations and 
	classical communications (LOCC). That is, for any given LOCC map $\Phi$ we have
	\be	\nonumber
	E[\Phi( \rho^{AB}) ]\leq E(\rho^{AB}),\quad\forall\;\rho^{AB}\in\mS^{AB}.
	\ee
	Moreover, convex measures of entanglement that do not increase \emph{on average}
	under LOCC are called entanglement monotones~\cite{Vidal2000}.

	Let $E$ be a measure of entanglement on bipartite states.  
	The entanglement of formation $E_F$ associated  with $E$ is defined by
	\begin{align}
	E_F(\rho^{AB})\equiv\min\sum_{j=1}^{n}p_jE(|\psi_j\lr\psi_j|^{AB}),\label{eofmin}
	\end{align}
	where the minimum is taken over all pure state decompositions of $\rho^{AB}=\sum_{j=1}^{n}p_j|\psi_j\lr\psi_j|^{AB}$.
	That is, $E_F$ is the convex roof extension  of $E$.
	Vidal~\cite[Theorem 2]{Vidal2000} showed that $E_F$ above is an entanglement monotone on mixed bipartite states  
	if the following concavity condition holds. For a pure state $|\psi\rangle^{AB}\in\mH^{AB}$, $\rho^A={\rm
		Tr}_B|\psi\rangle\langle\psi|^{AB}$, define the function $h:\mS^{A}\rightarrow\mathbb{R}_+$ by
	\be\label{h}
	h( \rho^A) \equiv E( |\psi\lr\psi|^{AB}).
	\ee 
	Note that since $E$ is invariant under local unitaries we must have
	\be\label{uinvariant}
	h( U\rho^AU^\dag) =h( \rho^A)
	\ee 
	for any unitary operator $U$ acting on $\mH^A$. If $h$ is also concave, i.e.
	\be \label{concave}
	h[\lambda\rho_1+(1-\lambda)\rho_2]\geq\lambda h(\rho_1)+(1-\lambda)h(\rho_2)
	\ee
	for any states $\rho_1$, $\rho_2$, and any
	$0\leq\lambda\leq1$, then $E_F$ as defined in~\eqref{eofmin} is an entanglement monotone.

	%\begin{theorem}
	\vspace{1mm}
	\noindent{\it \textbf{Theorem}.~Let $E$ be an entanglement monotone for which $h$, as defined in~\eqref{h}, 
		is strictly concave; i.e., $h$ satisfies~\eqref{concave} with strict inequality
	whenever $\rho_1\neq\rho_2$, and $0<\lambda<1$. Let also $E_F$ be as in~\eqref{eofmin}. Then,
	\begin{enumerate}
	\item If  $\rho^{ABC}=|\psi\lr\psi|^{ABC}$ is pure and~\eqref{cond} holds then 
	$\mH^B$ has a subspace isomorphic to $\mH^{B_1}\otimes\mH^{B_2}$ and up to local unitary on system $B_1B_2$,
	\be\label{product}
	|\psi\ra^{ABC}=|\phi\ra^{AB_1}|\eta\ra^{B_2C}\,,
	\ee
	where $|\phi\ra^{AB_1}\in\mH^{AB_1}$ and $|\eta\ra^{B_2C}\in\mH^{B_2C}$ are pure states.
	In particular, $\rho^{AC}$ is a product state (and consequently $E(\rho^{AC})=0$), so that $E$ is monogamous on pure tripartite states.
	\item If  $\rho^{ABC}$ is a mixed tripartite state and 
	$E_F(\rho^{A|BC})=E_F(\rho^{AB})$,
	then 
	\be
	\rho^{ABC}=\sum_{x}p_x|\psi_x\ra\la\psi_x|^{ABC},
	\ee 
	where $\{p_x\}$ is some probability distribution, and for each $x$ the Hilbert space
	$\mH^B$ has a subspace isomorphic to $\mH^{B_1^{(x)}}\otimes\mH^{B_2^{(x)}}$ such that 
	up to local unitary on system $B$, each pure state $|\psi_x\ra^{ABC}$ is  given by
	\be
	|\psi_x\ra^{ABC}=|\phi_x\ra^{AB_1^{(x)}}|\eta_x\ra^{B_2^{(x)}C}\,.
	\ee
	 In particular, the marginal state $\rho^{AC}$ is separable so that $E_F$ is monogamous (on mixed tripartite states).
	\end{enumerate}
    }
	%\end{theorem}
	\begin{remark}
	The condition that $E$ in the theorem above is an entanglement monotone 
	can be replaced with a weaker condition that the measure of entanglement $E$ satisfies $E\leq E_F$ on all bipartite density matrices.
	This is due to the fact that both Theorem 4 and Corollary 5 in Ref.~\cite{GG} are still true if we assume only that
	$E$ satisfies $E\leq E_F$ and $E$ is not necessarily 
	an entanglement monotone.
	\end{remark}

\begin{remark}
	Part 1 of the Theorem indicates that, if $E$ is an
	entanglement monotone with $h$ is strictly concave, 
	then for pure state $|\psi\ra^{ABC}$, $E(\rho^{AB})= E_F(\rho^{AB})$ provided that $E(|\psi\ra^{A|BC})=E_F(\rho^{AB})$, 
	that is, $E(|\psi\ra^{A|BC})=E(\rho^{AB})$ is equivalent to $E(|\psi\ra^{A|BC})=E_F(\rho^{AB})$ in such a case.
	Corollary 5 in Ref.~\cite{GG}
	proved only that if $E(|\psi\ra^{A|BC})=E(\rho^{AB})$,
	then $E(\rho^{AB})= E_F(\rho^{AB})$, but it is unknown whether $E(|\psi\ra^{A|BC})=E_F(\rho^{AB})$
	can imply $E(\rho^{AB})= E_F(\rho^{AB})$ since in general we have only $E(\rho^{AB})\leq E_F(\rho^{AB})$ 
	(e.g., for the negativity $N$, we have $N\leq N_F$).
\end{remark}

	\begin{proof}
	{\it Part 1.} In Ref.~\cite{GG} it was shown that if~\eqref{cond} holds for a 
	pure tripartite state $\rho^{ABC}\equiv |\psi\lr\psi|^{ABC}$ then 
	all pure state decompositions of $\rho^{AB}$ must have the same average entanglement.
	Let $\rho^{AB}=\sum_{j=1}^{n}p_j|\psi_j\lr\psi_j|^{AB}$ 
	be an arbitrary pure state decomposition of $\rho^{AB}$ with $n=\rank(\rho^{AB})$. Then, 
	\be\nonumber
	E( \rho^{AB}) \leq E_F( \rho^{AB}) =\sum_{j=1}^{n}p_jE( |\psi_j\lr\psi_j|^{AB}), 	
	\ee
	where the inequality follows from the convexity of $E$, and the equality 
	holds since all pure state decompositions of $\rho^{AB}$ have the same average entanglement. 
	Moreover, since $E_F$ is an entanglement monotone, we must have 
	\be\nonumber
	E_F( \rho^{AB}) \leq E(  |\psi\lr\psi|^{A|BC}) =h(\rho^A).
	\ee
	Therefore, denoting by $\rho_j^A\equiv\tr_B|\psi_j\lr\psi_j|^{AB}$, 
	we conclude that if~\eqref{cond} holds then we must have
	\be\nonumber
	\sum_{j=1}^{n}p_jh( \rho_j^A) =h( \rho^A).
	\ee
	Given that $\rho^A=\sum_{j=1}^{n}p_j\rho_j^A$ and $h$ is strictly concave we must have
	\be
	\rho_{j}^A=\rho^A,\quad j=1,...,n.
	\ee
	Set $r\equiv\rank(\rho^A)\leq \dim\mH^B$, and let $\mH^{B_1}$ be 
	an $r$-dimensional subspace of $\mH^B$ such that there exists a pure 
	state $|\phi\ra^{AB_1}\in\mH^{AB_1}$ with marginal on part $A$ being $\rho^A$. 
	Since all the reduced density matrices of $\{|\psi_j\ra^{AB}\}$ have 
	the same marginal on system $A$ they must be related via local isometry on Bob's side, 
	to a purification $|\phi\ra^{AB_1}$ of $\rho^A$. Therefore, there exists 
	isometries $\{V_j^{B_1\to B}\}$ such that
	\be\label{psij}
	|\psi_j\ra^{AB}=(I^A\otimes V_j^{B_1\to B})|\phi\ra^{AB_1},\quad j=1,...,n.
	\ee
	Now, let $\rho^{AB}=\sum_{k=1}^{n}q_k|\phi_k\lr\phi_k|^{AB}$ be another 
	pure state decomposition of $\rho$ with the same number of elements $n$. 
	For the same reasons leading to~\eqref{psij}, there exists isometries $W_k^{B_1\to B}$ such that
	\be\nonumber%\label{phik}
	|\phi_k\ra^{AB}=(I^A\otimes W_k^{B_1\to B})|\phi\ra^{AB_1},\quad k=1,...,n.
	\ee
	On the other hand, since both decompositions $\{p_j,|\psi_j\ra^{AB}\}$ and $\{q_k,|\phi_k\ra^{AB}\}$ 
	correspond to the same density matrix $\rho^{AB}$, 
	they must be related by a unitary matrix $U=(u_{kj})$ in the following way:
	\ba\nonumber
	\sqrt{q_k}|\phi_k\ra^{AB}&=\sum_{j=1}^{n}u_{kj}\sqrt{p_j}|\psi_j\ra^{AB}\\
	&=\left(I^A\otimes \sum_{j=1}^{n}u_{kj}\sqrt{p_j}V_j^{B_1\to B}\right)|\phi\ra^{AB_1}.	
	\ea
	Denoting by
	\be\nonumber
	X_k^{B_1\to B}\equiv\frac{1}{\sqrt{q_k}}\sum_{j=1}^{n}u_{kj}\sqrt{p_j}V_j^{B_1\to B},\quad k=1,...,n\;,
	\ee
	we have
	\be\nonumber
	(I^A\otimes X_k^{B_1\to B})|\phi\ra^{AB_1}=(I^A\otimes W_k^{B_1\to B})|\phi\ra^{AB_1}\;.
	\ee
	Now, multiplying both sides of the equation above by $(\rho^A)^{-1/2}$ 
	(the inverse is understood to be on the support of $\rho^A$), we get
	\be\nonumber
	(I^A\otimes X_k^{B_1\to B})|\phi_+\ra^{AB_1}=(I^A\otimes W_k^{B_1\to B})|\phi_+\ra^{AB_1},
	\ee
	and we therefore conclude that $X_k^{B_1\to B}=W_k^{B_1\to B}$. 
	This means that $X_k^{B_1\to B}$ is an isometry for any choice of unitary matrix $U=(u_{kj})$.
	But since $U=(u_{jk})$ is an arbitrary unitary matrix, 
	we can take its first row, $\{u_{1j}\}_j$ to be an arbitrary normalized vector. 
	Hence, we conclude that any linear combination of the isometric matrices 
	$\{V_j^{B_1\to B}\}$ is proportional to an isometric matrix. 
	We now discuss the consequence of this property on the form of $\{V_j^{B_1\to B}\}$.

	The isometries $V_j^{B_1\to B}$ can be expressed as
	\be
	V_j^{B_1\to B}=\sum_{k}|v_{jk}\lr k|\,,
	\ee
	where $\{|k\ra\}$ is an orthonormal basis of $\mH^{B_1}$, and for each $j$, 
	$\{|v_{kj}\ra\}_{k}$ are some orthonormal vectors in $\mH^B$. 
	Consider the arbitrary linear combination $\sum_jc_jV_j$. It can be expressed as
	\be\nonumber
	\sum_{k,j}c_j|v_{kj}\lr k|\equiv \sum_{k}|u_k\lr k|,\quad |u_{k}\ra\equiv\sum_{j}c_j|v_{kj}\ra.
	\ee
	Therefore, $\sum_jc_jV_j$ is proportional to an isometry if and only if for 
	all $k\neq k'$, $\la u_{k'}|u_{k}\ra=0$ and $\|u_k\|=\|u_{k'}\|$. Observe that
	\be\nonumber
	\la u_{k'}|u_{k}\ra=\sum_{j,j'}c_j{c}_{j'}^*\la v_{k'j'}|v_{kj}\ra.
	\ee
	Now, for a fixed $k$ and $k'$, the above equation can be viewed as an 
	inner product between a vector $\bv_{kk'}$, whose components are $\la v_{k'j'}|v_{kj}\ra$, 
	and a vector $\bar{\bc}\otimes\bc$, whose components are $c_{j}^*c_{j'}$. 
	Since $\bv_{kk'}$ is orthogonal to any vector of the form $\bar{\bc}\otimes\bc$ whenever $k\neq k'$, 
	it must be equal to the zero vector. We therefore conclude that
	\be\nonumber
	\la v_{k'j'}|v_{kj}\ra=0,\quad k\neq k'
	\ee
	and 
	\be\nonumber
	\la v_{kj'}|v_{kj}\ra=d_{jj'}\neq 1
	\ee
	for some $d_{jj'}$ which are independent of $k$.
	Note that we can assume with no loss of generality that
	$\rho^{AB}=\sum_jp_j|\psi_j\ra\la\psi_j|^{AB}$ is the spectral decomposition of $\rho^{AB}$.
	In such a case,
	we have
	\begin{eqnarray*}\nonumber
		&&\la\psi_{j'}|\psi_j\ra^{AB}\\
		&=&\la \phi|^{AB_1}\left (I^A\otimes \sum_{k,k'}|k'\ra\la v_{k'j'}|v_{kj}\ra\la k|\right) |\phi\ra^{AB_1}\\
		&=&\la \phi|^{AB_1}\left (I^A\otimes \sum_{k}|k\ra\la v_{kj'}|v_{kj}\ra\la k|\right) |\phi\ra^{AB_1}\\
		&=&\la \phi|^{AB_1}\left [I^A\otimes d_{jj'} \left( \sum_{k}|k\ra\la k|\right) \right] |\phi\ra^{AB_1}\\
		&=&\la \phi|^{AB_1}\left (I^A\otimes  d_{jj'}I^{B_1}\right) |\phi\ra^{AB_1}
		=\delta_{jj'}
	\end{eqnarray*}
	 holds for any given $k$.
	 We now obtain that 
	 \be
	 \la v_{k'j'}|v_{kj}\ra=\delta_{jj'}\delta_{kk'}.
	 \ee
	Denote $\mK\equiv\spa\{|v_{kj}\ra\}\subset\mH^B$. Then, the equation above 
	implies that $\mK\cong\mH^{B_1}\otimes\mH^{B_2}$ for some subspace $\mH^{B_2}$ of $\mH^B$, and in particular,
	there exists a unitary matrix, $U^B$, relating the basis elements $\{|v_{kj}\ra\}$ 
	of $\mK$ with the basis elements $\{|k\ra^{B_1}|j\ra^{B_2}\}$ of $\mH^{B_1}\otimes\mH^{B_2}$. 
	We therefore conclude that
	\begin{align}
	V_j^{B_1\to B}=\sum_{k}|v_{kj}\lr k|&=U^B\left(\sum_{k}|k\lr k|^{B_1}\otimes |j\ra^{B_2}\right)\nonumber\\
	&=U^B\left( I^{B_1}\otimes|j\ra^{B_2}\right).\nonumber
	\end{align}
 Hence, 	
	\be\label{psij2}
	|\psi_j\ra^{AB}=( I^A\otimes U^B) |\phi\ra^{AB_1}|j\ra^{B_2},\quad j=1,...,n.
	\ee
	Therefore, 
	\be\nonumber
	\rho^{AB}=( I^A\otimes U^B) (|\phi\lr\phi|^{AB_1}\otimes \sigma^{B_2})( I^A\otimes U^B) ^\dag,
	\ee
	 where $\sigma^{B_2}$ is some density matrix given by $\sigma^{B_2}=\sum_jp_j|j\lr j|^{B_2}$.
	 Therefore, $\rho^{AB}$ has a purification of the form
	 \be
	 (I^{A}\otimes U^B\otimes I^{C})|\phi\ra^{AB_1}|\eta\ra^{B_2C},
	 \ee
	 where $|\eta\ra^{B_2C}\equiv \sum_jp_j|j\ra^{B_2}|j\ra^{C}$. Since all purifications 
	 of $\rho^{AB}$ in $\mH^{ABC}$ are related via a local unitary on $C$, we conclude 
	 that $|\psi\ra^{ABC}$ has the form~\eqref{product} up to local unitary. 
	 Therefore $\rho^{AC}$ is a product state
	 and consequently $E(\rho^{AC})=0$.
	This completes the proof of part 1. 
	
	{\it Proof of Part 2:} Let $\{p_x,\;|\psi_x\ra^{ABC}\}$ be the optimal pure 
	state decomposition of $\rho^{ABC}=\sum_xp_x|\psi_x\lr\psi_x|^{ABC}$ such that 
	\be
	E_F(\rho^{A|BC})=\sum_xp_xE(|\psi_x\ra^{A|BC}).
	\ee
	Denote by $\rho_x^{AB}\equiv\tr_C(|\psi_x\lr\psi_x|^{ABC})$ and note that 
	$$
	\rho^{AB}\equiv \tr_C(\rho^{ABC})=\sum_xp_x \rho_x^{AB}.
	$$
	Hence, if $E_F(\rho^{A|BC})=E_F(\rho^{AB})$ then
	\be\nonumber
	\sum_xp_xE(|\psi_x\ra^{A|BC})
	=E_F( \rho^{AB}) \leq \sum_xp_x E_F( \rho_x^{AB}),	
	\ee
	where we used the convexity of $E_F$. On the other hand, since $E_F$ is a 
	measure of entanglement for each $x$ we have $E(|\psi_x\ra^{A|BC})\geq E_F(\rho_x^{AB})$. 
	Combining this with the equation above we conclude that
	\be
	E(|\psi_x\ra^{A|BC})= E_F(\rho_x^{AB})\,,\quad\forall\;x\;.
	\ee
	Therefore, the rest of the proof of part 2 follows from part 1 of the theorem.		
		\end{proof}

Note that if system $B$ in the second part of the theorem above has dimension not greater than 3, 
then we must have for each $x$ that either $\mH^{B_1^{(x)}}$ or ${\mH}^{B_{2}^{(x)}}$ 
are one dimensional. We therefore get the following corollary.

	\vspace{1mm}
	%\begin{cor}
	\noindent{\it \textbf{Corollary}.~Using the same notations as in the theorem above, 
		if $E_F(\rho^{A|BC})=E_F(\rho^{AB})$ and $\dim\mH^B\leq 3$ then $\rho^{ABC}$ is bi-separable, 
		and in particular it admits the form
		\be
		\rho^{ABC}=t\sigma^{A|BC}+(1-t)\gamma^{AB|C},
		\ee
		where $\sigma^{A|BC}$ is $A|BC$ separable, $\gamma^{AB|C}$
		is $AB|C$ separable, and $t\in[0,1]$.
		In particular, if $\rho^{ABC}=|\psi\ra\la\psi|^{ABC}$ is a pure state, then
		$|\psi\ra^{ABC}$ has the form $|\phi\ra^{AB}|\eta\ra^{C}$ or $|\phi\ra^{A}|\eta\ra^{BC}$.}	
	%\end{cor}
	\vspace{1mm}

	At last we discuss the strict concavity of the entanglement measures so far. 
	Many operational measures of entanglement such as the relative entropy of entanglement, 
	entanglement cost, and distillable entanglement, all reduce on a bipartite pure 
	state to the entropy of entanglement given in terms of the von Neumann entropy 
	of the reduced state, $H(\rho)\equiv-\tr(\rho\log\rho)$. 
	The von Neumann entropy is known to be strictly concave~\cite{Wehrl} 
	and therefore they are all monogamous on pure tripartite states.	
	The first part of the theorem above generalizes a similar result 
	that was proved in Ref.~\cite[the disentangling theorem]{Hehuan} 
	for the special case in which $E$ is taken to be the negativity. 
	It demonstrates that many measures of entanglement are monogamous on pure tripartite states, 
	while their convex roof extensions as defined in~\eqref{eofmin} 
	are monogamous even on mixed tripartite states.

	Any function that can be expressed as
	\be
	H_g(\rho)=\tr[g(\rho)]=\sum_jg(p_j)\,,
	\ee
	where $p_j$ are the eigenvalues of $\rho$ is strictly concave if $g''(p)<0$ for all $0<p<1$. 
	This includes the quantum Tsallis $q$-entropy~\cite{Tsallis,Landsberg}
	with $q>0$. In particular, the linear entropy (or the Tsallis $2$-entropy) 
	is strictly concave, and therefore the tangle %~\cite{Coffman} 
	is a monogamous measure of entanglement since it is defined in terms of the convex roof extension.
	Another important example is the R\'enyi $\alpha$-entropy~\cite{Greenberger,Renyi,Dur}. 
	For the R\'enyi parameter
	$\alpha\in[0,1]$ the R\'enyi entropies are strictly concave (see, e.g., Ref.~\cite{Vidal2000}), 
	but in general, for $\alpha>1$ the R\'enyi entropies are not even concave (although they are Schur-concave). 
	To the authors' knowledge, except for this case of R\'enyi $\alpha$-entropy of entanglement with $\alpha>1$, 
	all other measures of entanglement that have been studied intensively in literature, correspond 
	on pure bipartite state to strict concave functions of the reduced density matrix. 
	These include the negativity, tangle, concurrence (see the Appendix), 
	$G$-concurrence, and the Tsallis entropy of entanglement.

In conclusion, we showed that many measures of entanglement, 
such as the entanglement of formation, that were believed not to be monogamous 
(irrespective of the specific monogamy relation~\cite{Lan16}), 
are in fact monogamous according to a new definition of monogamy without 
inequalities that we put forward in Ref.~\cite{GG}. This new definition 
is equivalent to the quantitative inequality~\eqref{power}, but with a key 
difference that the exponent factor $\alpha$ can depend on the underlying dimension. 
Therefore, the results presented here support this non-universal 
(i.e., dimension dependent) definition of monogamy.  
The fact that so many important measures of entanglement are 
not universally monogamous~\cite{Lan16} may give the impression 
that monogamy of entanglement cannot be attributed to entanglement 
itself but rather is a property of the particular measure that 
is used to quantify entanglement. Furthermore, as was shown in Ref.~\cite{Lan16}, 
measures of entanglement cannot be simultaneously faithful 
(as defined in Ref.~\cite{Lan16}) and universally monogamous. 
Here we avoided all these issues by adopting a new definition 
of monogamy that allows for non-universal monogamy relations, 
while at the same time maintaining a quantitative way 
[as in~\eqref{power}] to express monogamy relations.

While we were not able to show that \emph{all} measures of entanglement 
are monogamous (according to our definition), we are also not aware 
of any continuous measures of entanglement that are not monogamous. 
It may be the case that all continuous measures of entanglement 
are monogamous, which will support our assertion that monogamy is 
a property of entanglement and not of some particular functions 
quantifying entanglement. Moreover, many important measures of 
entanglement, are not defined in terms of convex roof extensions. 
For such measures, our theorem does not provide any information 
regarding their monogamy on mixed tripartite states. 
One example of that is the negativity.
Our theorem implies that the convex roof extended negativity 
is monogamous but we do not know if the negativity itself is monogamous.

	\begin{acknowledgements}	
		The authors are very grateful to the referees for their 
		constructive suggestions. 
		Y.G was supported by the Natural Science Foundation of Shanxi Province
		under Grant No. 201701D121001, the National Natural
		Science Foundation of China under Grant No. 11301312, 
		 and the Program for the Outstanding Innovative 
		 Teams of Higher Learning Institutions of Shanxi.
		G.G.'s research was supported by the Natural Sciences 
		and Engineering Research Council of Canada (NSERC).	
	\end{acknowledgements}
	
	%\nocite{*}
	%\bibliographystyle{apsrev4-1}
	%\bibliography{gy0705}% Produces the bibliography via BibTeX.
	%\appendix*
	 	
		\appendix*
	\section{Constructing monogamous measures of entanglement from other monogamous measures}

	For any entanglement monotone $E$, and any monotonically increasing 
	function $g:\mbb{R}_{+}\to\mbb{R}_+$ with the property that $g(x)=0$ iff $x=0$, denote
	\be
	E_g( \rho^{AB}) \equiv\min\sum_{j}p_jg[E(|\psi_j\lr\psi_j|^{AB})],
	\ee
	where the minimum is taken over all pure state decompositions of 
	$\rho^{AB}=\sum_{j=1}^{n}p_j|\psi_j\lr\psi_j|^{AB}$. 
	Observe that if in addition $g$ is convex then we must have
	\be
	E_g( \rho^{AB}) \geq g[E( \rho^{AB})].
	\ee
	Therefore, if $E_g$ is an entanglement measure and is 
	monogamous on pure tripartite states then $E$ is also 
	monogamous on pure tripartite states. To see why, 
	note that if $E(\psi^{A|BC})=E(\rho^{AB})$ we also have
	\bea
	&&E_g( |\psi\ra^{A|BC}) =g[ E( |\psi\ra^{A|BC})]\nonumber\\
	&=&g[ E( \rho^{AB})]\leq E_g( \rho^{AB}). 
	\eea
	But since $E_g$ is a measure of entanglement we 
	must have $E_g(|\psi\ra^{A|BC})\geq E_g(\rho^{AB})$ so that 
	we get $E_g(|\psi\ra^{A|BC})=E_g(\rho^{AB})$. Since we assume 
	here that $E_g$ is monogamous, thus $E_g(\rho^{AC})=0$, 
	which implies that $E(\rho^{AC})=0$.
	As a simple example of this, consider the function 
	$g(x)=x^2$ and take $E=C$ be the concurrence as defined 
	in Ref.~\cite{Rungta2003pra}. Then,  $E_g=C^2$ is the 
	tangle which is monogamous (it is given in terms of the 
	linear entropy, which is strictly concave). 
	Hence, the above analysis implies that the concurrence $C$ is also monogamous.

\end{document}